\documentclass[twocolumn,twocolappendix]{aastex701}
\usepackage{graphicx}
\usepackage{amsmath}
\usepackage{CJK}

\begin{document}
\begin{CJK*}{UTF8}{gbsn}

\title{Magnetar X-ray flares as elementary reconnection events in a self-organized critical magnetosphere}
\shortauthors{Xie et al.}

\correspondingauthor{Sheng-Lun Xie, Yi Zhao, Wen-Long Zhang}
\email{xiesl@jgsu.edu.cn, yizhao@dzu.edu.cn, wlzhang@pmo.ac.cn}

\author[0000-0001-9217-7070]{Sheng-Lun Xie (谢升伦)}
\affiliation{Institute for Astronomy and Astrophysics, Jinggangshan University, Ji'an, Jiangxi 343009, China}
\affiliation{Key Laboratory of Energy Conversion Optoelectronic Functional Materials of Jiangxi Education Institutes, School of Mathematics and Physics, Jinggangshan University, Ji'an, Jiangxi 343009, China}
\email{xiesl@jgsu.edu.cn}

\author[0000-0002-4636-0293]{Yi Zhao (赵一)}
\affiliation{School of Computer and Information, Dezhou University, Dezhou 253023, People's Republic of China}
\email{yizhao@dzu.edu.cn}

\author[0009-0008-6247-0645]{Wen-Long Zhang (张文龙)}
\affiliation{Purple Mountain Observatory, Chinese Academy of Sciences, Nanjing 210023, China}
\affiliation{School of Astronomy and Space Sciences, University of Science and Technology of China, Hefei 230026, China}
\email{wlzhang@pmo.ac.cn}

\author[0009-0008-8053-2985]{Chen-Wei Wang (王晨巍)}
\affiliation{Department of Astronomy, Tsinghua University, Beijing 100084, China}
\email{cwwang@tsinghua.edu.cn}

\author[0000-0002-1067-1911]{Yun-Wei Yu (俞云伟)}
\affiliation{Institute of Astrophysics, Central China Normal University, Wuhan 430079, China}
\email{yuyw@ccnu.edu.cn}

\author[0000-0002-4771-7653]{Shao-Lin Xiong (熊少林)}
\affiliation{State Key Laboratory of Particle Astrophysics, Institute of High Energy Physics, Chinese Academy of Sciences, 19B Yuquan Road, Beijing 100049, China}
\email{xiongsl@ihep.ac.cn}

\begin{abstract}
Magnetar bursts are commonly attributed to magnetic energy release, yet the physically relevant dissipation unit has remained uncertain. Here we argue that this unit is not the burst as conventionally defined, but the short substructures within bursts, namely magnetar X-ray flares (MXFs). Using 789 MXFs from SGR~J1935+2154, we show that their waiting times separate into two components, interpreted as reflecting clustered intra-episode events and longer inter-episode intervals, that their proxy emission measure and temperature exhibit distinct rising and declining trends, and that their energy release scale (ERS) proxy distribution has a high-end slope in the same statistical regime as solar and stellar flares. Combined with the previously measured MXF duration slope, these results identify MXFs as a three-dimensional fractal-diffusive self-organized critical (FD-SOC) system with $\beta_{\rm diff}\simeq2$, implying reconnection avalanches that expand linearly through an active magnetosphere. We infer that magnetar active episodes are global unloading states assembled from many elementary flares. Magnetar X-ray flares are therefore the physically relevant units for magnetar outburst statistics and for connecting magnetar high-energy activity to other impulsive phenomena.
\end{abstract}

\keywords{
\uat{Magnetars}{992} ---
\uat{Magnetohydrodynamics}{1964} ---
\uat{Astrostatistics}{1882}
}

\section{Introduction}\label{sec:introduction}

Magnetars are neutron stars powered by magnetic fields of $10^{14}$--$10^{15}$ G and are best known for their impulsive X-ray and soft $\gamma$-ray activity \citep{Woods2006csxs,Kaspi2010PNAS,Mereghetti2013BrJPh,Olausen2014,KaspiBeloborodov2017ARAA,NegroYounes2024FrASS}. Their high-energy transients are conventionally grouped into ordinary short bursts, intermediate bursts or flares, and giant flares according to duration and energetics \citep{Olive2004ApJ,Woods2006csxs,Mereghetti2008AARv,Israel2008ApJ}. Yet the boundaries between these classes are not sharp, and magnetar activity spans a hierarchy of timescales from sub-second bursts to burst storms and long-term outbursts \citep{NegroYounes2024FrASS,Kaneko2021ApJ}. The conventional taxonomy provides a useful observational description, while the relation between these classes and the underlying dissipation process remains uncertain.

This ambiguity matters because magnetar burst statistics already resemble those of self-organized critical (SOC) systems. Power-law-like distributions of burst energy and waiting time have long suggested avalanche-type activity analogous to earthquakes or solar flares \citep{Cheng1995Natur,Gogus1999ApJ,Aschwanden2016SSRv,Xie2024ApJ}. Similar statistical behavior across physically distinct systems, such as gamma-ray bursts \citep[GRBs;][]{2013NatPh...9..465W,2023ApJS..265...56L,2026_SOC_07A} and fast radio bursts \citep[FRBs;][]{2017JCAP_Wang,2020MNRAS_Cheng,2021ApJ...920..153W}, however, also illustrates that SOC phenomenology alone does not identify the underlying dissipation process. Temporal statistics can characterize how events are organized, while additional observables tied to plasma heating and the energy release scale are needed to constrain where the observed dissipation occurs.

\begin{figure}
    \centering
    \includegraphics[width=1.0\linewidth]{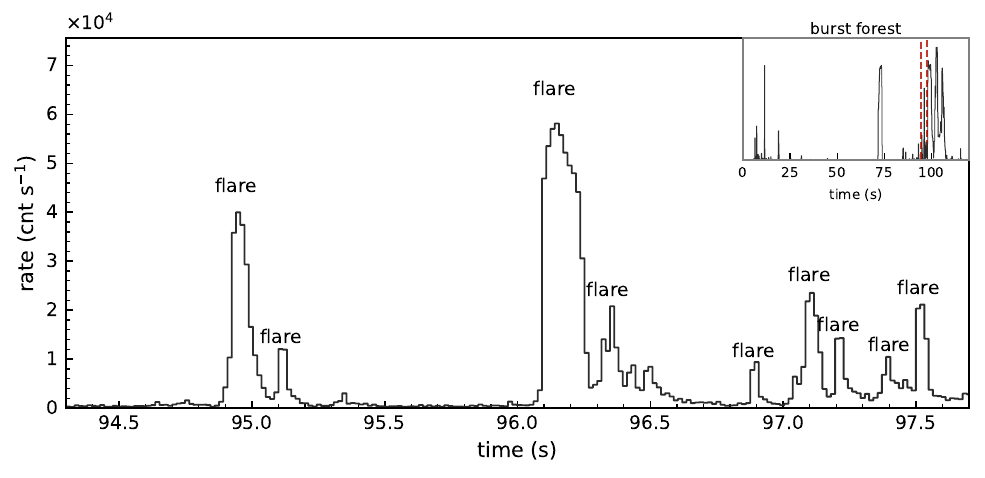}
    \caption{
    Burst-forest example from SGR~J1935+2154 observed with Fermi/GBM on 2020 April 27 at 18:31:30 UTC.
    In this work, the individual peaks are identified as magnetar X-ray flares, whereas the broader multi-peaked activity may be conventionally classified as bursts.
    }
    \label{fig:mxf_schematic}
\end{figure}

An additional complication is the choice of the observational unit on which these statistics are constructed. A single burst can combine multiple sub-peaks and multiple stages of one active episode, so statistical analyses performed on whole bursts may mix elementary and composite events. High time-resolution observations show that magnetar bursts are often composed of short-lived substructures rather than single homogeneous pulses. In practice, burst classification often relies on heuristic temporal criteria, such as whether neighboring peaks are separated by intervals comparable to or longer than the spin period. For magnetars with spin periods on the order of seconds, such a prescription is already ambiguous, and in burst forests it becomes even harder to apply in a physically meaningful way, as illustrated in Figure~\ref{fig:mxf_schematic}. In this paper, we refer to these short-lived substructures as magnetar X-ray flares (MXFs).

Previous work has shown that the temporal statistics of MXFs exhibit SOC-like scaling behavior \citep{Xiao2024ApJS,Xiao2025ApJ}. Building on these findings, we investigate whether ordinary MXFs can be identified with elementary reconnection events by combining their temporal organization with thermodynamic and energetic scaling, and whether magnetar active episodes can be understood as larger magnetospheric states assembled from them.

\section{Observations and Data Reduction}\label{sec:observations}

Discovered in 2014 \citep{Stamatikos2014gcn}, SGR~J1935+2154 is a Galactic magnetar that has since been monitored for more than 8 yr by multiple telescopes \citep[e.g.,][]{Israel2016MNRAS,Younes2020ApJ,Ridnaia2021NatAs,Borghese2022MNRAS,Cai2022b,Rehan2023apj,Xie2025ApJS}. It has attracted particular attention as a source of fast radio bursts \citep[e.g.,][]{Bochenek2020nat,CHIMEFRB2020nat,Li2021NatAs}.

We analyze a sample of 789 MXFs from SGR~J1935+2154, drawn from its active episodes \citep{Lin2020apj,Lin2020apjl,Xie2022mnras,Rehan2023apj,Rehan2024ApJ,Rehan2025ApJS,Xie2025ApJS}. We adopt the pulse peak times and durations listed in Table~2 of \citet{Xiao2024ApJS}. Time-integrated spectral analysis is performed using data from the Gravitational-wave high-energy Electromagnetic Counterpart All-sky Monitor \citep[GECAM;][]{LiWen2021,Xiao2022mnras} and the Fermi Gamma-ray Burst Monitor \citep[GBM;][]{Meegan2009apj}. For both instruments, we select detectors with detector-to-source angles $\leqslant 50^\circ$ in order to ensure good signal-to-noise ratios \citep{Collazzi2015apjs,Lin2020apj}. We use GECAM event-by-event (EVT) data and GBM time-tagged event (TTE) data to construct time-integrated spectra for individual MXFs.

For each MXF, the source spectrum is extracted over an interval whose total length is half of the catalogued pulse duration, defined by \citet{Xiao2024ApJS} as the time between the two valleys bounding the pulse.
The interval extends before and after the peak with a time ratio of 1:3, with the aim of concentrating on the main pulse emission and reducing contributions from neighboring pulses and background.
Background intervals of a few seconds before and after the burst are identified using the Bayesian blocks algorithm \citep{Scargle2013ApJ}. The background is modeled with a first-order polynomial based on these intervals. The resulting background spectrum is then used in the spectral fitting procedure.
Since the dominant emission of magnetar bursts is concentrated below $\sim 100$ keV, all fits are performed over the 8--200 keV energy range.

Because individual MXFs are short and photon-limited, the spectral evidence must be extracted with deliberately simple models.
Spectral fitting is carried out by minimizing the PG-statistic, which combines Poisson statistics for the total counts in the source interval with Gaussian uncertainties in the background estimate \citep{Cash1979ApJ,Arnaud1996ASPC}. 
Spectral channels are grouped using optimal binning \citep{Kaastra2016AA}, with a minimum of one total count per bin.
We checked alternative spectral binning schemes and found no appreciable changes in the main spectral results.

We consider three phenomenological models commonly used for magnetar-burst spectroscopy: optically thin thermal-bremsstrahlung (OTTB), single blackbody (BB), and power law (PL). Their mathematical forms are given in Appendix~\ref{sec:app_spectral_analysis}.
For each MXF, all candidate models are fitted and compared using the Akaike Information Criterion (AIC) and Bayesian Information Criterion (BIC), and the preferred model is identified by the minimum value \citep{Liddle2007MNRAS}.
Under the AIC and BIC test, 504, 213, and 72 events statistically favor OTTB, BB, and PL, respectively. These models are used as phenomenological descriptions rather than unique identifications of the radiation mechanism.
These fits provide a common basis for constructing comparable spectral and energetic proxies across the MXF sample, allowing us to test whether the decomposed events exhibit coherent flare-like scaling relations.

\section{Temporal Organization of MXFs}\label{sec:temporal}

Waiting times provide the first diagnostic of whether these substructures occur as independent events or as clustered building blocks of larger active states.

The waiting time between successive flares detected within the same observation window is defined as $\Delta t=t_i-t_{i-1}$, where $t_i$ is the peak time of the $i$th event. Because the waiting time of the first event in each observation window is undefined, only intervals between successive flares within the same window are retained, yielding 524 waiting times from the full set of 789 MXFs. This is a global analysis rather than one restricted to flares within individual bursts.

\begin{figure}
    \centering
    \includegraphics[width=1.0\linewidth]{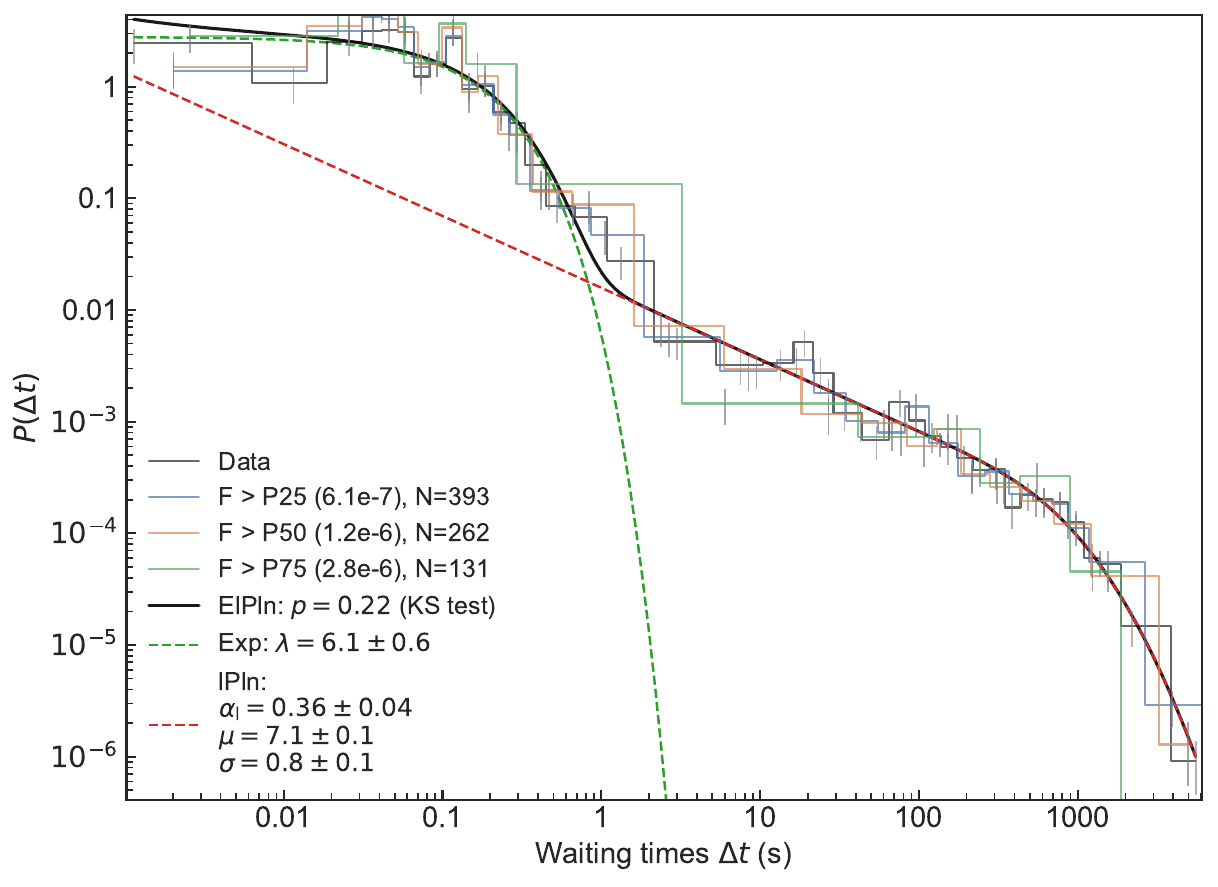}
    \caption{
    Waiting time probability distribution of MXFs from SGR~J1935+2154.
    Black points show the full sample, while colored points correspond to subsamples selected by flux thresholds in units of $\rm erg~cm^{-2}~s^{-1}$.
    The solid black curve shows the best-fitting ElPln model, consisting of an exponential component (green dashed line) and a left-Pareto lognormal component (red dashed line).
    The short waiting time regime is dominated by the exponential component, whereas the longer waiting time regime follows a power-law behavior characterized by a power-law slope $P(\Delta t)\propto\Delta t^{-0.64}$.}
    \label{fig:waiting_time_pdf_ElPln}
\end{figure}

Figure~\ref{fig:waiting_time_pdf_ElPln} shows that the MXF waiting time distribution is not described by a single timescale or a single scale-free law. It contains a quasi-flat short-timescale plateau and an extended long-time tail, and the same structure is recovered in subsamples selected by different flux thresholds. 
We initially fitted a double-Pareto lognormal (dPln) distribution to the longer intervals, $\Delta t>2~{\rm s}$. The fitted right-Pareto index is approximately 24, motivating an effective reduction to the left-Pareto lognormal (lPln) limiting form \citep{Reed2004CSTM,Xie2024ApJ}. Combining this component with an exponential distribution for the short intervals, we introduce the Exponential-left-Pareto lognormal (ElPln) mixture,
\begin{equation}\label{eq:elpln}
P_{\rm ElPln}(\Delta t) = C \cdot \lambda \exp(-\lambda \Delta t) + (1-C)\,P_{\rm lPln}(\Delta t),
\end{equation}
where $C\in[0,1]$ gives the mixture weight of the exponential component with rate parameter $\lambda$ and the left-Pareto lognormal (lPln) distribution \citep{Reed2004CSTM,Xie2024ApJ},
\begin{equation}\label{eq:lpln}
\begin{split}
P_{\rm lPln}(\Delta t) ={}& \alpha_{\rm l}\,(\Delta t)^{\alpha_{\rm l}-1}
\,A(-\alpha_{\rm l},\mu,\sigma)\\
&{}\times\Big[1-\Phi\!\left(z\right)\Big],\\
z={}&\frac{\ln(\Delta t)-\mu+\alpha_{\rm l}\,\sigma^2}{\sigma},
\end{split}
\end{equation}
where $A(\theta,\mu,\sigma)=\exp(\theta\mu+\theta^2\sigma^2/2)$, $\Phi(x)$ denotes the cumulative distribution function (CDF) of the standard normal distribution, $(\mu,\sigma)$ describes the lognormal core of the waiting time distribution, and $\alpha_{\rm l}$ is the Pareto index governing the small-$\Delta t$ scaling of the CDF, corresponding to $P(\Delta t) \sim \Delta t^{-(1-\alpha_{\rm l})} = \Delta t^{-\alpha_{\Delta t}}$ where $\alpha_{\Delta t}$ is the slope for waiting times.

The best-fitting ElPln model separates the distribution into an exponential component that dominates at $\Delta t \lesssim 2~{\rm s}$ and an lPln component that governs the longer intervals. Separate one-sample KS tests show that the short waiting times ($\Delta t<1~{\rm s}$) are consistent with an exponential distribution ($D_{\rm KS}=0.07$, $p=0.11$) and that the long waiting times ($\Delta t>2~{\rm s}$) are consistent with a double-Pareto lognormal distribution ($D_{\rm KS}=0.07$, $p=0.13$) \citep{Xie2024ApJ}. The short component is consistent with rapid flare occurrence within individual active episodes. In an avalanche picture, this component reflects the cadence of elementary reconnection events once a local magnetospheric region has been activated. 

At longer waiting times, the fitted Pareto index ($\alpha_{\rm l}=0.36\pm0.04$) gives $P(\Delta t)\propto\Delta t^{-0.64}$, much shallower than the waiting-time slopes commonly reported for earthquake aftershocks or solar flares \citep{Utsu1995,Bak2002PRL,Wheatland1998ApJ,Aschwanden2011book}. That shallow tail is difficult to reconcile with the slowly driven picture of independent avalanches discussed by \citet{Aschwanden2011book}, in which the waiting-time slope is expected to track the duration slope. It instead points to a higher-level component associated with larger events or active episodes, in which bursts of local activity are separated by longer intervals between episodes. The long-time part therefore requires additional physical interpretation beyond the local cadence of elementary reconnection events. Solar-flare studies show that an approximately constant local occurrence rate and its variation over longer timescales can produce exponential and broad waiting-time distributions, respectively \citep{Wheatland2000ApJL}. This provides a physical motivation for the mixture, whose specific functional form follows the fitting and simplification described above. The association of its components with intervals within and between active episodes is inferred from the global distribution.

Undetected weak or overlapping flares can merge consecutive intervals and increase the observed waiting times. As an additional check, we selected events above flux thresholds of $6.1\times10^{-7}$, $1.2\times10^{-6}$, and $2.8\times10^{-6}~{\rm erg\,cm^{-2}\,s^{-1}}$. All three samples yield power-law exponents of approximately $-0.6$, consistent with $-0.64\pm0.04$ for the full sample. The shallow power-law segment persists under these cuts.

Finite observation windows also affect the distribution. For a stationary renewal process observed through an independent continuous window of length $T_{\rm obs}$, the distribution of fully observed intervals obeys $p_{\rm obs}(\Delta t)\propto(1-\Delta t/T_{\rm obs})p_{\rm true}(\Delta t)$ for $0<\Delta t<T_{\rm obs}$ \citep{KivelaPorter2015PhRvE}. Longer intervals are therefore increasingly likely to extend beyond the window boundaries, limiting the information available at the longest waiting times. Future coordinated multi-instrument observations with comparable or well-characterized sensitivities can extend the effective continuous coverage and reduce temporal gaps, providing stronger constraints on the long waiting-time distribution.

A two-component waiting-time structure is known in other avalanche-type systems \citep{Wheatland1998ApJ,Corral2004PhRvL,Wang2023ApJ,Zhang2026arXiv260819713}, but here its physical implication is especially important. MXFs are organized in time as clustered sub-flares inside active episodes, not as isolated bursts. Waiting times alone do not yet identify the dissipation site, but they already argue that the burst should be decomposed before one asks what the elementary release unit actually is.

\section{Thermodynamic Scaling of MXFs}\label{sec:thermodynamic}\label{sec:scaling}

For the OTTB-favored subset of 504 MXFs, we derive an emission-measure proxy $\tilde{\rm EM}$ to track the amount of emitting plasma in a form that is comparable, at the level of scaling, with solar and stellar flare studies.

The proxy emission measure is constructed by relating the fitted OTTB spectrum to the standard optically thin thermal-bremsstrahlung emissivity,
\begin{equation}
\epsilon_{\nu}^{\rm ff}\propto Z^2 n_e n_i\,T^{-1/2}\exp\left(-\frac{h\nu}{k_{\rm B}T}\right)\bar g_{\rm ff}(T,\nu),
\end{equation}
where $n_e$ and $n_i$ are the electron and ion number densities, $Z$ is the ionic charge, and $\bar g_{\rm ff}$ is the velocity-averaged Gaunt factor \citep{RybickiLightman1986book}. For an optically thin source at distance $D_{\rm L}$, the corresponding photon number spectrum is
\begin{equation}
\begin{split}
N(E)\propto{}& \frac{1}{4\pi D_{\rm L}^2}
\left(\int Z^2 \bar g_{\rm ff} n_e n_i\,dV\right)\\
&{}\times(k_{\rm B}T)^{-1/2}E^{-1}
\exp\left(-\frac{E}{k_{\rm B}T}\right).
\end{split}
\end{equation}
The OTTB normalization therefore scales as $K\propto {\rm EM}_{\rm ff}(k_{\rm B}T)^{-1/2}$, where ${\rm EM}_{\rm ff}\equiv \int Z^2 \bar g_{\rm ff} n_e n_i\,dV$. We therefore define
\begin{equation}
\tilde{\rm EM}=K\,(k_{\rm B}T)^{1/2},
\end{equation}
as the corresponding emission measure proxy. 
For comparisons among events, we approximate the composition-dependent factors and the relevant Gaunt-factor dependence as common normalization factors, so that variations in $\tilde{\rm EM}$ trace relative changes in the emitting plasma content within this approximation. Under this approximation, $\tilde{\rm EM}$ is proportional, up to an overall constant factor, to the usual emission measure used in solar- and stellar-flare studies \citep{Feldman1995ApJ,Aschwanden2008ApJ}.

\begin{figure}
    \centering
    \includegraphics[width=1\linewidth]{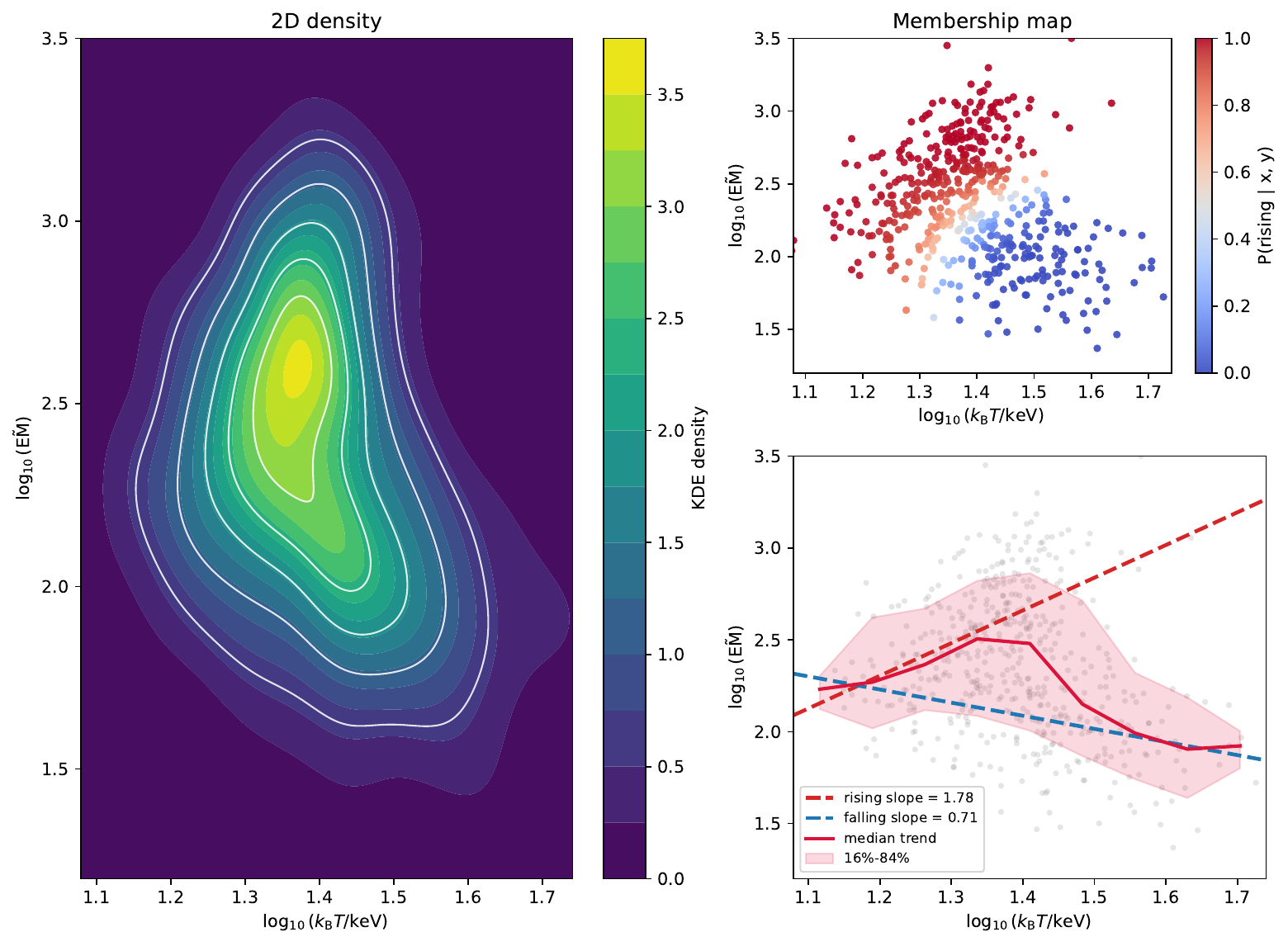}
    \caption{
    Distribution of the emission measure proxy $\tilde{\rm EM}$ as a function of temperature for MXFs statistically favoring the OTTB spectral model. 
    Left: two-dimensional kernel-density estimate in the $\log_{10}(k_{\rm B}T)$--$\log_{10}\tilde{\rm EM}$ plane. 
    Upper right: probabilistic membership map, where the color indicates the posterior probability that an event belongs to the rising branch, $P({\rm rising}\mid x,y)$. 
    Lower right: empirical binned median trend (solid red line) and the corresponding 16\%--84\% percentile range of the data (shaded region), together with the best-fitting rising and falling branches from the two-component mixture model (dashed lines). 
    }
    \label{fig:emT_mixture}
\end{figure}

Figure~\ref{fig:emT_mixture} shows that the OTTB-favored MXFs do not lie on a single thermodynamic sequence. Instead, they form two branches in the $\tilde{\rm EM}$--$T$ plane, one rising and one declining. The lower-right panel shows that this is a genuine population structure rather than a blurred single trend.

To quantify the apparent two-trend structure between temperature and emission measure, we describe the conditional distribution of $y=\log_{10}\tilde{\rm EM}$ at a given $x=\log_{10}(k_{\rm B}T/{\rm keV})$ as a smooth mixture of two linear relations in log--log space. 
Each component follows a Gaussian distribution around a linear trend $y \simeq a_i + b_i x$ ($i=1,2$) with an effective scatter $\Sigma_i$. 
The full conditional probability density can therefore be written as
\begin{equation}\label{eq:two_linear}
p(y\mid x)=\pi(x)\,{\cal N}_1 + \bigl[1-\pi(x)\bigr]\,{\cal N}_2 ,
\end{equation}
where ${\cal N}_i \equiv \mathcal{N}\!\left(y;\,a_i+b_i x,\,\Sigma_i^2\right)$.
Here $\Sigma_i^2 = s_i^2+\sigma_y^2$ combines the intrinsic dispersion $s_i$ and the measurement uncertainty $\sigma_y$. 
The mixing fraction is modeled as a logistic transition $\pi(x)=\left[1+\exp\!\left(-(x-x_{\rm b})/w\right)\right]^{-1}$, with transition location $x_{\rm b}=\log_{10}(k_{\rm B}T_{\rm b}/{\rm keV})$ and width $w$. 
This formulation allows the two correlations to coexist over a finite temperature range and naturally reproduces the observed hybrid trend.

The fit is performed on the unbinned flare sample by maximizing the likelihood over the full eight-parameter model, with multiple initializations used to reduce sensitivity to local extrema. The intrinsic scatters are parameterized in logarithmic form to enforce positivity, and the two components are relabeled after convergence according to slope, such that the component with the larger slope is identified as the rising branch, as plotted in Figure~\ref{fig:emT_mixture}. The fitted branches have slopes $b_{\rm rise}=1.78\pm0.23$ and $b_{\rm fall}=-0.71\pm0.21$, respectively. Parameter uncertainties are estimated from the local curvature of the likelihood and further checked with posterior sampling.
PIT and coverage diagnostics are consistent with a calibrated conditional distribution for the adopted two-component model.
We further tested the persistence of the two-branch structure using relaxed selection thresholds of $\Delta{\rm BIC}\leqslant2$, 6, and 10. These thresholds yield expanded sets of 615, 677, and 697 MXFs, respectively (see Table~\ref{tab:em_bic_sensitivity}), in each of which both rising and declining branches persist. The corresponding AIC thresholds yield identical event selections. Subsequent analyses and discussion continue to use the set with $\Delta{\rm BIC}\leqslant0$.

\begin{table}
    \caption{
    Fitted $\tilde{\rm EM}$--$T$ branch slopes under different spectral-model selection thresholds.
    The first column gives the selection threshold for $\Delta{\rm BIC}={\rm BIC}_{\rm OTTB}-\min({\rm BIC}_{\rm BB},{\rm BIC}_{\rm PL})$.
    $N$ is the number of selected MXFs. The fitted slopes $b_{\rm rise}$ and $b_{\rm fall}$ are reported with their $1\sigma$ statistical uncertainties, estimated from the local curvature of the likelihood.
    AIC yields identical event selections.
    }
    \label{tab:em_bic_sensitivity}
    \begin{tabular}{cccc}
        \hline
        $\Delta{\rm BIC}$ & $N$ & $b_{\rm rise}$ & $b_{\rm fall}$ \\
        \hline
        $\leqslant0$  & 504 & $1.78\pm0.23$ & $-0.71\pm0.21$ \\
        $\leqslant2$  & 615 & $1.08\pm0.20$ & $-0.67\pm0.18$ \\
        $\leqslant6$  & 677 & $1.06\pm0.19$ & $-0.73\pm0.18$ \\
        $\leqslant10$ & 697 & $1.04\pm0.19$ & $-0.74\pm0.17$ \\
        \hline
    \end{tabular}
\end{table}

\section{Energy Release Scale of MXFs}\label{sec:energy_release_scale}

A geometric area can be inferred only for the BB-favored subset and therefore cannot organize the full sample. We instead introduce a model-agnostic 
\textbf{energy release scale (ERS)}, 
denoted by $S_{\gamma}$, as
\begin{equation}\label{eq:ers}
S_{\gamma} = F\,E_{\rm c}^{-\gamma},
\end{equation}
where $F$ is the flare energy flux, $\gamma$ is a phenomenological scaling index, and $E_{\rm c}$ is a \textbf{characteristic spectral energy} 
defined by
\begin{equation}\label{eq:Ec}
E_{\rm c} = \frac{\int E\,N(E)\,{\rm d}E}{\int N(E)\,{\rm d}E}
\end{equation} 
Here, $N(E)$ is the photon number spectrum at photon energy $E$, so $E_{\rm c}$ provides a measure of the average photon energy.
In the special case of optically thick thermal emission, where the luminosity is proportional to $R^2 T^{4}$ and $E_{\rm c}\propto T$, one expects $\gamma \approx 4$, so that $S_{\gamma}$ scales with emitting area. More generally, $S_{\gamma}$ is a phenomenological scale coordinate that can be applied across spectral classes without assuming one common radiation model.

The ERS analysis uses the full set of 789 MXFs, with one best-fitting spectral model selected for each flare by the BIC. For every event we first compute the characteristic energy $E_{\rm c}$ entering Eq.~\ref{eq:Ec} by numerically integrating the preferred photon spectrum over the fitted energy range, 8--200~keV. Because the normalization cancels in Eq.~\ref{eq:Ec}, $E_{\rm c}$ depends only on the spectral shape parameter: $k_{\rm B}T$ for the BB and OTTB models, and the photon index $\alpha$ for the PL model. We then evaluate Eq.~\ref{eq:ers} using the measured energy fluxes.
The distribution fit itself is performed on the unbinned raw $S_\gamma$ sample.

The probability distribution of $S_{\gamma}$ exhibits a mixed structure consisting of a lognormal-like core and a power-law tail, which can be well described by the right-Pareto lognormal (rPln) distribution \citep{Reed2004CSTM,Xie2024ApJ},
\begin{equation}\label{eq:rpln}
\begin{split}
P_{\rm rPln}(S_{\gamma}) ={}& \alpha_{\rm r}\, S_{\gamma}^{-\alpha_{\rm r}-1}
\,A(\alpha_{\rm r},\mu,\sigma)\,\Phi(z),\\
z={}&\frac{\ln (S_{\gamma})-\mu-\alpha_{\rm r}\sigma^{2}}{\sigma},
\end{split}
\end{equation}
where $\alpha_{\rm r}$ is the right-tail Pareto index, corresponding to $P(S_{\gamma}) \sim S_{\gamma}^{-(1+\alpha_{\rm r})} = S_{\gamma}^{-\alpha_{S}}$.

To fit the ERS distribution, we use the rPln form given by Eq.~\ref{eq:rpln} and optimize the scaling index $\gamma$ jointly with the distribution parameters. For numerical stability, the location parameter is written as $\mu=\mathrm{median}[\ln S_\gamma]+\Delta\mu$, so the fitted parameter vector becomes $(\gamma,\Delta\mu,\sigma,\alpha_{\rm r})$. The unbinned log-likelihood is constructed by evaluating the rPln density at each transformed scale $S_{\gamma,i}=F_iE_{{\rm c},i}^{-\gamma}$ and including the Jacobian associated with this change of variable. The preferred parameter values are obtained from MLE fitting, and the uncertainties quoted in the main text are estimated from the local curvature of the likelihood surface.

\begin{figure}
    \centering
    \includegraphics[width=0.45\textwidth]{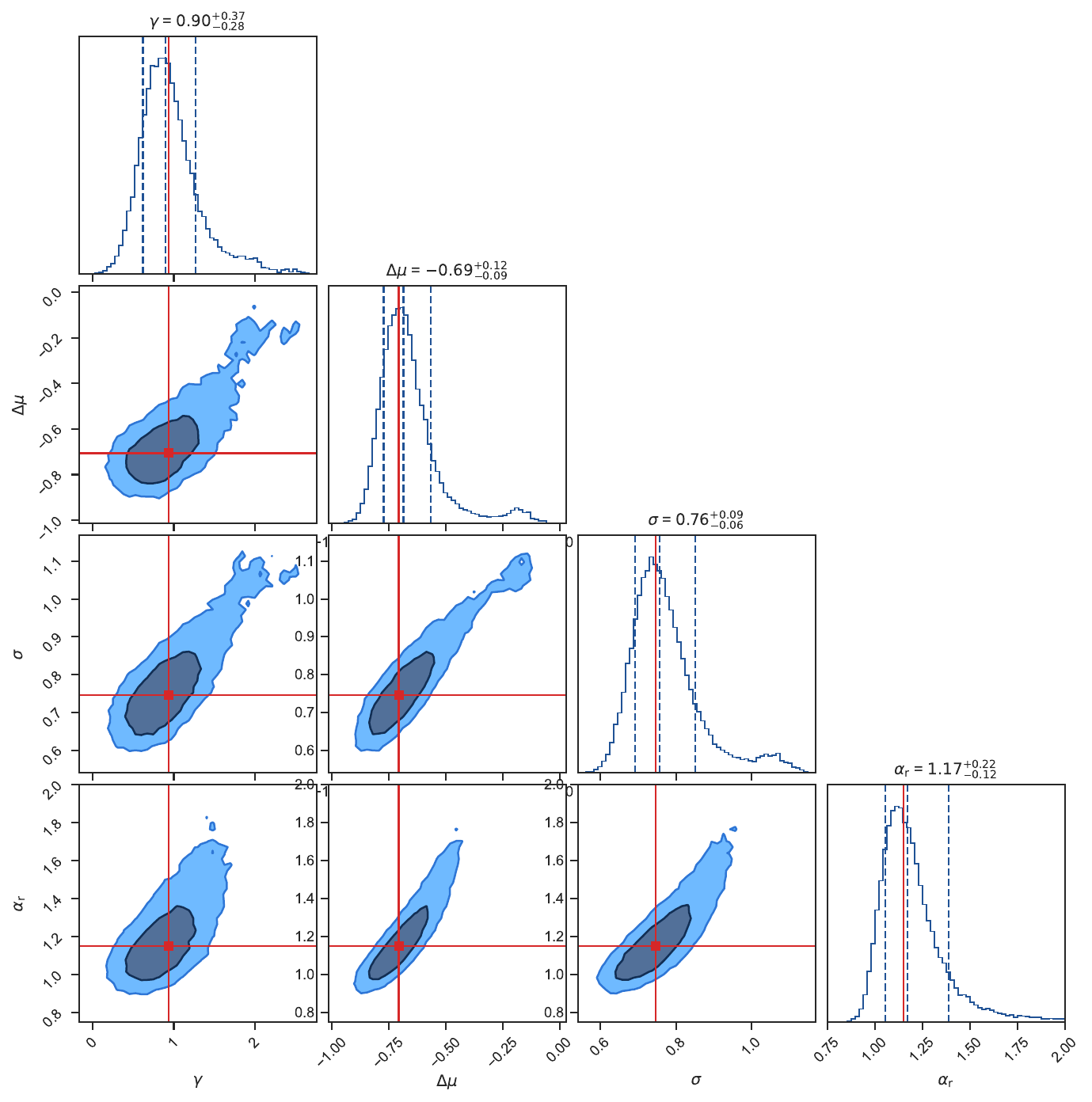}
    \caption{
    Posterior distributions of the ERS model parameters from the joint rPln fit. 
    The fitted parameters are the scaling index $\gamma$, the offset $\Delta\mu$ that defines the rPln location parameter, the lognormal width $\sigma$, and the right-Pareto index $\alpha_{\rm r}$. 
    Vertical red lines indicate the MLE result, and the dashed lines mark the 16th, 50th, and 84th percentiles.
    }
    \label{fig:ers_corner}
\end{figure}

We additionally perform posterior sampling as a consistency check, adopting broad bounded priors for the rPln parameters together with a weak Gaussian prior on $\gamma$. The posterior constraints are consistent with the MLE result, and the corner plot in Figure~\ref{fig:ers_corner} shows correlations of $\gamma$ with the location offset $\Delta\mu$, width $\sigma$, and tail index $\alpha_{\rm r}$. Changing $\gamma$ shifts and reshapes the ERS distribution, allowing partial compensation by these parameters. The results shown here are obtained from the joint fit. This joint-likelihood framework allows the preferred scale index to be inferred directly from the full flare ensemble without binning the data or fixing $\gamma$ a priori.

A joint likelihood fit gives $\gamma = 0.94 \pm 0.29$. This is far below the reference value for ideal blackbody area scaling $\gamma \approx 4$, and is consistent with unity. At $\gamma=1$, Eqs.~\ref{eq:ers} and \ref{eq:Ec} give $S_1=F/E_{\rm c}$, which is proportional to the photon number flux in the fitted energy band, up to the fixed conversion between energy units. We interpret this scaling in terms of non-equilibrium dissipation in transient magnetospheric plasma, consistent with the dominance of OTTB-favored events and with the waiting-time evidence for clustered local activity \citep{Duncan1992ApJ,Masada2010PASJ,Yu2011ApJ,Yu2013ApJ}.
This behavior is difficult to identify in conventional burst-level analyses because a single burst can mix multiple flares from different sites or from different stages of one avalanche episode. Once the sample is decomposed into MXFs, the scale distribution becomes interpretable as the statistics of elementary release units rather than as the average of composite burst phenomenology.

\begin{figure*}
    \centering
    \gridline{
        \fig{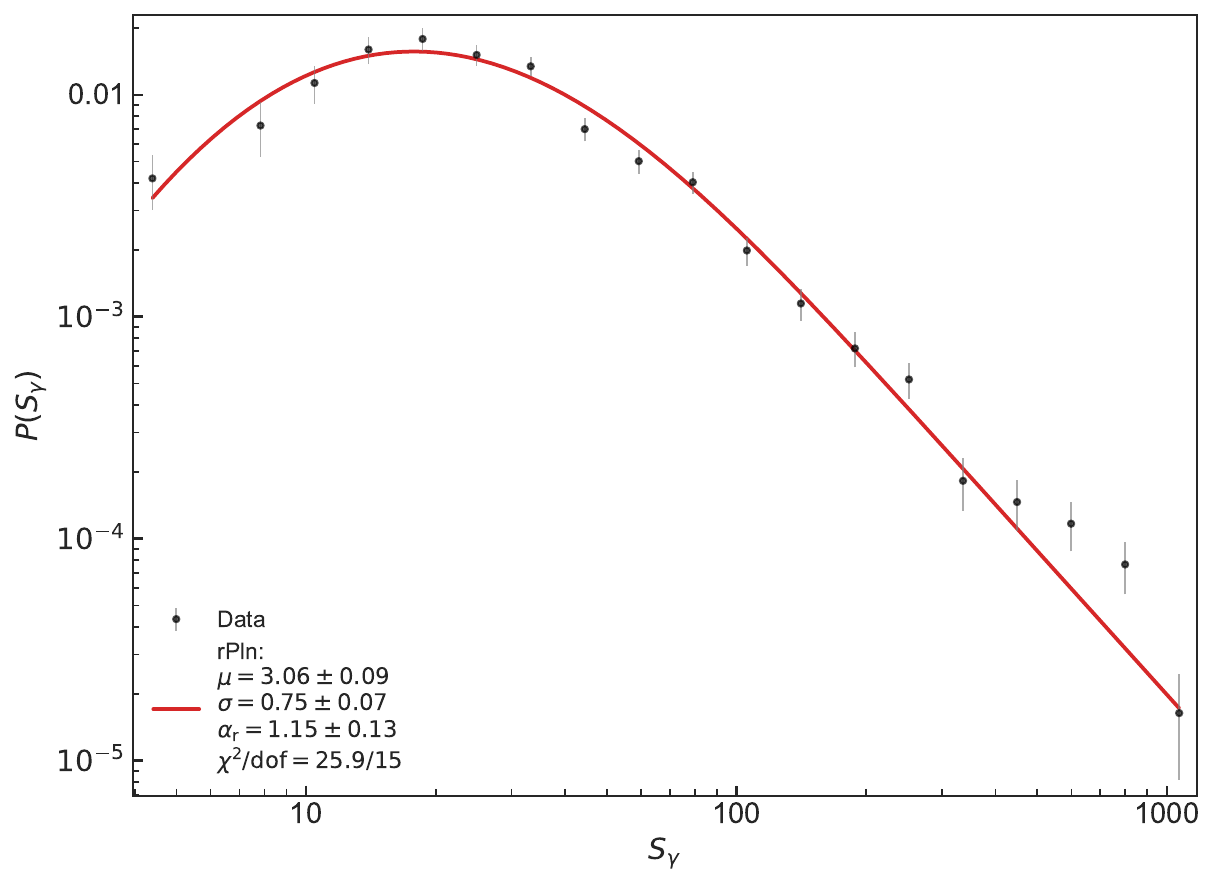}{0.48\textwidth}{(a)}
        \fig{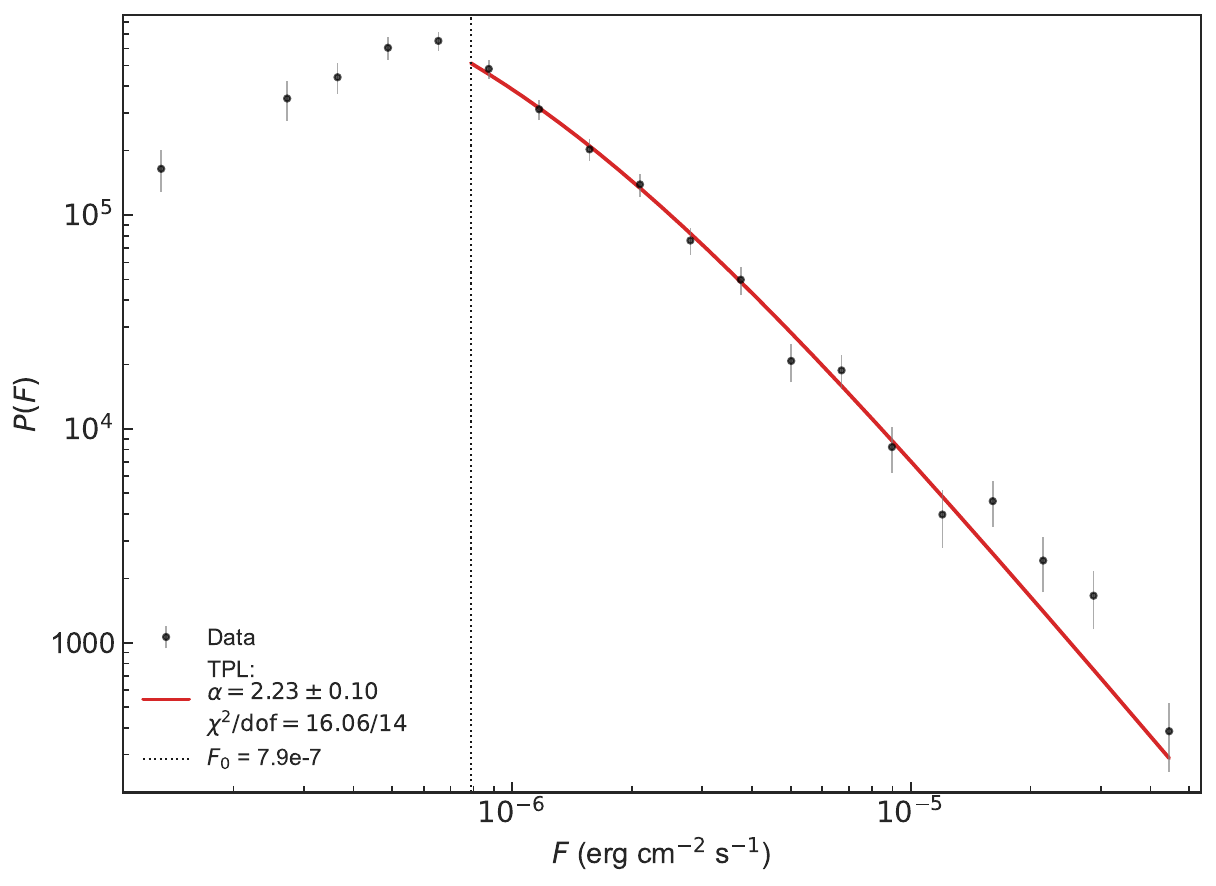}{0.48\textwidth}{(b)}
    }
    \caption{
    Distributions of the ERS and energy flux of MXFs.
    (a) Probability distribution of the phenomenological ERS proxy.
    Black points show $S_{\gamma}$ evaluated at $\gamma=0.94$, and the solid red curve shows the best-fitting right-Pareto lognormal (rPln) model.
    (b) Probability distribution of the energy flux.
    Gray points with error bars show the observed distribution, and the red solid curve shows the best-fitting thresholded power-law (TPL) model.
    The vertical dotted line marks the threshold flux
    $F_0=7.9\times10^{-7}\ {\rm erg\ cm^{-2}\ s^{-1}}$.
    The best-fitting high-flux slope is $\alpha_F=2.23\pm0.10$.
    }
    \label{fig:scale_flux}
\end{figure*}

Figure~\ref{fig:scale_flux}(a) shows a lognormal-like ERS core plus a clear power-law tail at large scales, with $\alpha_{S}=1+\alpha_{\rm r}=2.15\pm0.13$. This is the cleanest geometric-statistical bridge to flare physics in other systems. Although $S_{\gamma}$ is not a literal area, its high-end slope falls in the same regime as the area distributions of solar flares, whose compiled indices cluster near $\alpha_A\approx2$ \citep{Aschwanden2011book}. That is precisely the range expected for avalanche-like release in a three-dimensional SOC system, and it argues that MXFs trace the effective radiative scale of elementary magnetospheric flares.

As an independent energetic cross-check, we examine the MXF flux distribution. Figure~\ref{fig:scale_flux}(b) shows that it is well described by a thresholded power law (TPL) \citep{Aschwanden2015ApJ}. Above a characteristic threshold flux $F_0$, the distribution follows
\begin{equation}\label{eq:tpl}
P(F) \propto (F+F_0)^{-\alpha_F},
\end{equation}
where the threshold term captures the low-flux turnover from incompleteness and the high-flux tail captures the intrinsic scale-free part of the population.

The best-fitting slope is $\alpha_F = 2.23 \pm 0.10$ with $\chi^2/{\rm dof}=16.06/14$. The high-flux MXF population therefore follows a robust power law, and its slope is essentially identical to the values commonly reported for solar-flare flux distributions, which cluster near $\alpha_F\sim2$ \citep{Aschwanden2011book}. Together with the ERS tail, this places MXFs in the same energetic scaling regime as reconnection-driven flare populations in a critical SOC state.


\section{Discussion and Conclusions}\label{sec:discussion}

The central result of this work is that the temporal, thermodynamic, geometric, and energetic statistics of MXFs all converge on the same picture. The ERS tail with $\alpha_S = 2.15 \pm 0.13$, and the flux slope $\alpha_F = 2.23 \pm 0.10$ place MXFs in the same statistical regime as solar and stellar flares. The two-branch $\tilde{\rm EM}$--$T$ relation further allows the thermodynamic response of MXFs to be compared with the heating and plasma filling processes inferred in solar and stellar flares. When these results are combined with the MXF duration slope $\alpha_T \simeq 3$ \citep{Xiao2024ApJS,Xiao2025ApJ}, they support a specific physical interpretation of the MXF population as a three-dimensional fractal-diffusive self-organized critical (FD-SOC) system. The dominant dissipation is therefore more naturally attributed to magnetospheric flaring than to direct crustal energy release.

\subsection{Magnetospheric Avalanches and Activity Hierarchy}\label{subsec:discuss_avalanches}

Aschwanden's generalized macroscopic FD-SOC description has been applied to solar flares and soft gamma-ray repeaters; for a three-dimensional system, the previously published MXF duration slope $\alpha_T \simeq 3$ implies $\beta_{\rm diff} \simeq 2$ \citep{Aschwanden2014ApJ,Xiao2024ApJS,Xiao2025ApJ}. The avalanche front then expands linearly, with the activated-region scale $r(t) \propto t^{\beta_{\rm diff}/2} \simeq t$, through progressively larger magnetospheric structures once the source enters an active state. This scaling describes the growth of the activated region and provides a dynamical context for the thermodynamic trends discussed below. The preferred ERS scaling lies far below the blackbody expectation, and the large-scale tails remain in the flare regime. If starquakes participate, they are better understood as the process that loads or triggers the magnetosphere than as the site of the dominant observed dissipation.

The waiting-time structure gives this active state a concrete meaning. The short exponential-like component is the rapid cadence of elementary flares within a locally activated magnetosphere. The much shallower long-time component does not follow the slowly driven waiting-time picture of independent avalanches discussed by \citet{Aschwanden2011book} and is more naturally interpreted as a higher-level modulation associated with larger events or active episodes. The cutoff near $\Delta t \gtrsim 500$~s may arise if the disturbance expands through a closed magnetosphere of finite extent. 

This interpretation is aimed at the ordinary MXFs that dominate the flare statistics. It does not require every magnetar high-energy transient to belong to the same physical regime. Rare giant flares, and perhaps some intermediate flares, may involve large-scale restructuring, substantial plasma expansion, or trapped fireball formation and may therefore lie outside the regime probed by the ordinary MXF population \citep{Thompson2001ApJ,Gaensler2005Natur,Xie2026aApJ,Xie2026bApJ}. The broader implication is a hierarchical phenomenology of magnetar activity, in which the abundant low-energy events are elementary magnetospheric flares and the rare highest-energy events may access additional dynamics.

We note that the MXF statistics inherit selection effects from finite time resolution, detection thresholds, and pulse overlap in the adopted catalogue \citep{Xiao2024ApJS}. The association of individual peaks with elementary dissipation events also requires further assessment. Moreover, power-law indices across multiple observables do not uniquely identify a three-dimensional FD-SOC system, since power-law flare statistics also arise in cellular-automaton SOC models \citep{Lu1993ApJ} and non-SOC stochastic models of independent reconnection elements \citep{WheatlandCraig2003ApJ}. Future work should test the robustness of pulse identification and its impact on the inferred statistics, and examine model predictions for correlations between flare duration and radiative output to help distinguish these interpretations.

\subsection{Thermodynamic Comparison with Solar and Stellar Flares}\label{subsec:discuss_thermodynamics}

\citet{Shibata1999ApJ} derive a conditional EM--$T$ reference for solar and stellar flares from a reconnection-heated loop model. In this model, reconnection heating is balanced by thermal conduction; the conducted energy heats chromospheric plasma, and chromospheric evaporation supplies dense plasma to the reconnected loop. The evaporated loop plasma is confined when its gas pressure is comparable to the magnetic pressure. The peak-temperature, pressure, loop scale and EM relations are
\begin{equation}
\begin{aligned}
T &\propto B^{6/7} n_0^{-1/7} L^{2/7},\\
T &\sim \frac{B^2}{16\pi\: n k_{\rm B} },\\
{\rm EM} &\sim n^2 L^3.
\end{aligned}
\end{equation}
Here, $B$ is the local magnetic field, $L$ is the characteristic scale of the flare loop, $n_0$ is the ambient density before reconnection, and $n$ is the density of the hot loop plasma. The corresponding EM--$T$ scalings are
\begin{equation}
\begin{aligned}
{\rm EM} &\propto B^{-5} n_0^{3/2} \cdot T^{17/2},\\
 &\propto L^{5/3} n_0^{2/3} \cdot T^{8/3}.
\end{aligned}
\end{equation}
At fixed $B$ and $n_0$, the temperature exponent is $17/2$; at fixed $L$ and $n_0$, it is $8/3$. These conditional scalings provide reference slopes for a phenomenological comparison with the MXF population, without assuming identical loop physics in magnetar magnetospheres and the solar corona.

The rising branch resembles the familiar solar and stellar flare sequence in which stronger events heat more plasma and reach higher temperatures \citep{Feldman1995ApJ,Aschwanden2008ApJ}. Numerically, the measured rising-branch slope of $1.78\pm0.23$ is closer to the fixed-$L$ exponent $8/3$ than to the fixed-$B$ exponent $17/2$ \citep{Shibata1999ApJ}. This comparison links stronger heating with greater plasma filling or retention within a restricted range of emitting scales, while allowing the emitting scale, particle supply, and energy-transport conditions to vary among events.

The solar and stellar population studies cited above chiefly establish positive EM--$T$ sequences. Solar observations also reveal distinct thermal components within individual flares, including hot components associated with chromospheric evaporation and super-hot coronal components, with the latter often exhibiting high emission measures and electron densities \citep{Caspi2010ApJL,Caspi2014ApJ}. These results illustrate that different heating and plasma-filling conditions can coexist within solar flares, whereas the MXF branches describe trends across a population of events. This distinction motivates comparison of the underlying heating and filling processes while retaining the MXF branches as population-level observables.

The declining branch may reflect a different flare state, in which heating accompanies growth of the emitting region, limited particle filling, or particle loss, allowing the emission measure to fall as temperature rises. As a simplified physical description for approximately uniform density and fixed composition,
\begin{equation}
\begin{aligned}
{\rm EM} &\simeq \frac{N_e^2}{V_{\rm em}},\\
\frac{d\ln {\rm EM}}{dt} &= 2\frac{d\ln N_e}{dt}-\frac{d\ln V_{\rm em}}{dt},
\end{aligned}
\end{equation}
where $N_e$ is the total number of electrons in the emitting region and $V_{\rm em}$ is its effective radiating volume. The relation makes explicit how particle supply and retention compete with volume growth: sufficient filling can sustain or increase EM during heating, whereas faster volume growth, incomplete filling, or particle loss can lower it. Continued heating can therefore coexist with a higher temperature and a lower retained emission measure. The linear growth of the activated-region scale $r(t)$ provides a dynamical context for region development, while $V_{\rm em}$ describes the volume occupied by the radiating plasma. Different balances between heating, particle retention, and the growth of the emitting region could therefore contribute to the contrasting trends across the MXF population.

The solar-loop scaling rests on conductive transport, chromospheric evaporation, and a gas-pressure/magnetic-pressure relation. Magnetar conditions require a dedicated physical treatment because local magnetic fields, strong magnetization, relativistic reconnection, radiative transport, and pair creation can alter heating, confinement, and radiation \citep{Uzdensky2011SSRv}. The emitting geometry and particle source also require specification, as does the relation between fitted temperature and physical plasma temperature. One should note that $\tilde{\rm EM}$ is a proxy derived from time-integrated MXF spectra. These differences define the modeling task of establishing heating, particle supply, and radiation conditions appropriate to magnetar flares.

\subsection{Implications for FRB--X-ray Associations}\label{subsec:discuss_frb}

Decomposing bursts into MXFs therefore addresses more than nomenclature. Statistics organized by burst can combine multiple MXFs from different locations or stages of an active episode, obscuring the underlying magnetic release events. The first FRB--XRB association from SGR~J1935+2154 was established by the temporal correspondence between the two radio pulses of FRB~200428 and two narrow X-ray peaks \citep{CHIMEFRB2020nat,Bochenek2020nat,Li2021NatAs,Ridnaia2021NatAs}. \citet{Ge2023ApJ} decomposed the Insight-HXMT light curve into these two narrow peaks and broader burst components. The combined spectrum of the two narrow peaks has a photon index of $\sim 1.4$, which is softer than that of the spectrum fitted over the same time interval including both the narrow peaks and broader components ($\sim 0.9$). This comparison shows that a spectrum integrated over the burst can mix distinct emission components. Identifying the X-ray sub-pulses that coincide with the radio pulses therefore provides a more direct basis for characterizing the high-energy emission linked to the radio pulses. MXF decomposition specifies the release events and spectral properties compared across radio and X-ray observations, clarifying which X-ray component is paired with each radio pulse.

\citet{Wang2026MNRAS} report a further selective correspondence in MXB/FRB 221014, where the X-ray burst contains two narrow pulses, one temporally aligned with the main radio pulse. The waiting-time distribution of the MXFs studied here describes the temporal organization of high-energy release. If detectable radio counterparts arise from only a subset of MXFs, additional high-energy release events can occur between two radio detections. Radio waiting times would then depend both on the release sequence and on the conditions for radio production, propagation, and visibility. Differences in heating, particle supply, confinement, and losses may affect coherent radio production and escape. The EM--$T$ relation motivates tests of whether these conditions vary among MXFs. Recent statistical studies of repeating FRBs have likewise revealed multiscale temporal correlations and stable energy-release hierarchies across changing activity levels, providing a radio-domain analogy to the hierarchical organization inferred here \citep{Zhang2026arXiv260819713,Yang2026arXiv260818455}. MXF decomposition therefore ties radio/X-ray associations to individual release events and helps distinguish the organization of magnetic activity from the conditions governing its radiation.

\begin{acknowledgments}
This work was supported by the Shandong Provincial Natural Science Foundation (Grant No.~ZR2024QA181), the Scientific Research Fund of Dezhou University (Grant No.~2024xjrc142), and the National Natural Science Foundation of China (Grant No.~42605096).
\end{acknowledgments}

\facilities{Fermi (GBM), GECAM}

\software{
GECAMTools \citep{GECAMTools2024},
Fermi GBM Data Tools \citep{GbmDataTools},
elisa \citep{ELISA}
}

\appendix
\restartappendixnumbering
\section{Phenomenological Spectral Models}\label{sec:app_spectral_analysis}

We consider three phenomenological models commonly used for magnetar-burst spectroscopy: optically thin thermal-bremsstrahlung (OTTB), single blackbody (BB), and power law (PL). 
The OTTB model is written as
\begin{equation}
    N(E) = K\,E^{-1}\exp\left(-\frac{E}{k_{\rm B}T}\right),
\end{equation}
where $k_{\rm B}T$ is the plasma temperature in units of keV, and $K$ is the normalization in units of $\rm photons\,cm^{-2}\,s^{-1}\,keV^{-1}$ at 1 keV.

The BB model is written as
\begin{equation}
    N(E)=1.0344 \times 10^{-3} \cdot K\,\frac{E^2}{\exp(E/k_{\rm B}T)-1},
\end{equation}
where $k_{\rm B}T$ is the blackbody temperature in units of keV,
and $K=(R/D_{\rm L})^2$ is the normalization, with $R$ the apparent blackbody radius in units of km and $D_{\rm L}$ the source distance in units of 10 kpc.

The PL model is written as
\begin{equation}
    N(E) = K\,E^{-\alpha},
\end{equation}
where $\alpha$ is the photon index and $K$ is the normalization in units of $\rm photons\,cm^{-2}\,s^{-1}\,keV^{-1}$ at 1 keV.



\bibliography{main}{}
\bibliographystyle{aasjournalv7}

\end{CJK*}
\end{document}